\documentclass{IEEEtran}
\usepackage{multirow}
\usepackage{fixltx2e}
\usepackage{graphicx}
\usepackage{algpseudocode}
\usepackage{amssymb}
\usepackage{algorithm}
\usepackage{amsmath}
\newcommand{\subparagraph}{}
\usepackage[small,compact]{titlesec}
\usepackage{paralist}
\usepackage{multicol}
\usepackage{cite}
\usepackage[labelfont=bf, font={scriptsize,bf}]{caption}
\usepackage[utf8]{inputenc}
\usepackage{caption}
\usepackage{microtype}
\usepackage{subfig}
\usepackage[none]{hyphenat}
\usepackage{url}
\usepackage{fancybox}
\usepackage{tikz}
\usepackage{pgfplots}
\usepackage{float}
\algtext*{EndWhile}
\algtext*{EndIf}
\algtext*{EndFor}
\algtext*{End}
\algnewcommand{\LineComment}[1]{\State \(\triangleright\) #1}
\DeclareCaptionType{copyrightbox}
\def\scaleLarge{3cm}
\begin{document}
\sloppy

\def \sectionSpace{1mm}

\def\lrpc{\mbox{\sc CCLLRPC}}
\def\run{\mbox{\sc Run}}
\def\arun{\mbox{\sc ARun}}
\def\rems{\mbox{\sc Rem'}s}
\def\remsp{\mbox{\sc RemSP}}
\def\nremsp{\mbox{\sc CCLRemSP}}
\def\aremsp{\mbox{\sc ARemSP}}
\def\paremsp{\mbox{\sc PARemSP}}
\def\merge{\mbox{\sc Merge}}
\def\flatten{\mbox{\sc Flatten}}
\def\merger{\mbox{\sc Merger}}

\newcommand{\systemName}[0]{\emph{WikiTables}}
\newcommand{\systemNameHeader}[0]{WikiTables}
\newcommand{\columnRankingTaskName}[0]{\emph{relevant join}}
\newcommand{\tableSearchTaskName}[0]{\emph{table search}}
\newcommand{\columnRankingTaskNameHeader}[0]{Relevant Join}
\newcommand{\tableSearchTaskNameHeader}[0]{Table Search}
\newcommand*{\myfont}{\fontfamily{pzc}\selectfont}

\newcommand{\tableCorpus}[0]{\cal{T}}
\newcommand{\queryTable}[0]{\mathbf{T}_{q}}
\newcommand{\targetTable}[0]{\mathbf{T}_{t}}
\newcommand{\finalTable}[0]{\mathbf{T}_{f}}
\newcommand{\matchCorpus}[0]{\cal{M}}
\newcommand{\matchPerc}[0]{\emph{MatchPercent}}

\newcommand{\sourceCol}[0]{\mathbf{c}^{q}_{s}}
\newcommand{\matchedCol}[0]{\mathbf{c}^{t}_{m}}
\newcommand{\addableCol}[0]{\mathbf{c}^{t}_{c}}
\newcommand{\addedCol}[0]{\mathbf{c}^{t}_{a}}
\newcommand{\sarmaSys}[0]{\textsc{Rel\_Table}}
\newcommand{\ventisSys}[0]{\textsc{Table}}
\newcommand{\rjBaseline}[0]{\textsc{Base}}
\newcommand{\rjAblateSR}[0]{\emph{WikiTables$-$SR}}
\newcommand{\rjBest}[0]{\emph{WikiTables}}
\newcommand{\googft}[0]{\textsc{Google Fusion Tables}}
\newcommand{\venetisEtAl}[0]{Venetis et al.}
\newcommand{\sarmaEtAl}[0]{Sarma et al.}
\def\Section {\S}

\title{A New Parallel Algorithm for Two-Pass Connected Component Labeling} 

\author{
\IEEEauthorblockN{Siddharth Gupta, Diana Palsetia, Md. Mostofa Ali Patwary,
Ankit Agrawal, Alok Choudhary} \\
\IEEEauthorblockA{Department of Electrical Engineering \& Computer Science,
Northwestern University, Evanston, IL 60208, USA \\  
siddharth.bitsian@gmail.com, \{drp925,
mpatwary, ankitag, choudhar\}@eecs.northwestern.edu}}

\maketitle

\begin{abstract}
Connected Component Labeling (CCL) is an important step in pattern
recognition and image processing. It assigns labels to the pixels such that
adjacent pixels sharing the same features are assigned the same label.
Typically, CCL requires several passes over the data. We focus on two-pass technique where 
each pixel is given a provisional label in the first pass whereas an actual
label is assigned in the second pass.

We present a scalable parallel two-pass CCL algorithm, called \paremsp,\ 
which employs a scan strategy and the best union-find
technique called \remsp,\ which uses \rems\ algorithm for storing label
equivalence information of pixels in a 2-D image. In the first pass, we divide the image among threads and each
thread runs the scan phase along with \remsp\
simultaneously. In the second phase, we assign the final labels to the pixels. As \remsp\ is easily parallelizable, we use the parallel
version of \remsp\ for merging the pixels on the boundary. Our experiments
show the scalability of \paremsp\ achieving speedups up to $20.1$ using $24$
cores on shared memory architecture using {\em OpenMP} for an image of size
$465.20$ MB. We find that our proposed parallel algorithm achieves
linear scaling for a  large resolution fixed problem size as the number of
processing elements are increased. Additionally, the parallel algorithm does
not make use of any hardware specific routines, and thus is highly portable.
\end{abstract}

 \vspace{\sectionSpace}
\section{Introduction}

One of the most fundamental operations in pattern recognition is the labeling of connected components in a binary image.
Connected component labeling (CCL) is a procedure for assigning a unique label
to each object (or a connected component) in an image. Because these labels are key for other analytical procedures, connected component labeling is an 
indispensable part of most applications in pattern recognition and computer vision, such as fingerprint identification,
character recognition, automated inspection, target recognition, face identification, medical image analysis, and 
computer-aided diagnosis. In many cases, it is also one of the most time-consuming tasks among other 
pattern-recognition algorithms\cite{Alnuweiri1992_Parallel}. Therefore, connected component labeling continues 
to be an active area of research \cite{Gonzales_Digital,Agarwal2006_Efficient,Chang2004_Linear,Hayashi2001_Fast,
Hu2005_Fast,Knop1998_Parallel,Moga1997_Parallel,Wang2003_Parallel}.


There exist many algorithms for computing connected components in a given image. These algorithms are categorized into
mainly four groups 
\cite{Suzuki2003_Linear}
: $1)$ repeated pass
algorithms\cite{Haralick1981_Repeated,Hashizume1990_Algorithm}, $2)$ two-pass
algorithms\cite{gotoh1990high,gotoh1987component,komeichi1989video,lumia1983new,lumia1983new_c,naoi1995high,rosenfeld1970connectivity,
rosenfeld1966sequential,shirai1987labeling}
$3)$ Algorithms with hierarchical tree equivalent representations of the
data\cite{dillencourt1992general,gargantini1982separation,hecquard1991connected,samet1981connected,samet1985computing,samet1988efficient,
samet1986improved,tamminen1984efficient},
$4)$ parallel
algorithms\cite{bhattacharya1996connected,choudhary1994connected,
hirschberg1979computing,manohar1989connected,nassimi1980finding,olariu1993fast}.
The repeated pass algorithms perform repeated passes over an image in forward and backward raster directions alternately
to propagate the label equivalences until no labels change.
In two-pass algorithms, during the first pass, provisional labels are assigned to connected components;
the label equivalences are stored in a one-dimensional or a two-dimensional table array. After the first pass, the label 
equivalences are resolved by some search. This step is often performed by using
a search algorithm such as the union-find algorithm.
The results of resolving are generally stored in a one-dimensional table. During the second pass, the provisional labels are 
replaced by the smallest equivalent label using the table. Since the algorithm traverses image twice, these algorithms 
are called two-pass algorithms.
In algorithms that employ hierarchical tree structures i.e., n-ary tree such as binary-tree, quad-tree, octree, etc., the
label equivalences are resolved by using a search algorithm such as the union-find algorithm.
Lastly, the parallel algorithms have been developed for parallel machine models such as a mesh connected massively parallel processor.
However all these algorithms share one common step, known as scanning step in
which provisional label is given to each of the pixel depending on its neighbors.

In this paper we focus on two-pass CCL algorithms. The algorithm in \cite{Wu2009_LRPC} and \cite{He2012_ARun} are two
developed techniques for two-pass Connected Component Labeling.
The algorithm in \cite{Wu2009_LRPC}, which we refer to as \lrpc, uses a decision tree to assign provisional labels and an 
array-based union-find data structure
to store label equivalence information. However, the technique employed for
union-find, Link by Rank and Path Compression is not the best technique available \cite{Patwary2012_PARemSP}. 
The algorithm in \cite{He2012_ARun}, which we refer to as \arun, employs a special scan order over the data and three linear
arrays instead of the conventional union-find data structure. There
exists a parallel implementation of \arun\ on TILE\-64 many core
platform\cite{Chen2013_PARun}. According to the
experimental results given in \cite{Chen2013_PARun}, the parallel implementation is able to achieve a speedup of 10 on
32 processor units. 
This implementation is also not portable due to its implementation for specific hardware architecture.

\begin{table}[h]
\caption{Abbreviations used in the paper and their brief description }
\centering
\begin{tabular}{l p{5cm}} 
\hline\hline
Abbreviation & Description\\ [0.5ex] 
\hline 
CCL & Connected Component Labeling\\[1ex]
\arun\ & CCL algorithm suggested by \cite{He2012_ARun}\\[1ex]
\remsp\ & union-find technique proposed by {\em Rem}
\cite{Patwary2010_RemSP}\\[1ex]
\aremsp\ & CCL algorithm proposed in our paper using scan strategy of \arun\ and
\remsp\\[1ex]
\paremsp\ & Parallel implementation of \aremsp\ proposed in our paper\\[1ex]
\lrpc\ & CCL algorithm suggested by \cite{Wu2009_LRPC}\\[1ex]
 \nremsp\ & CCL algorithm proposed in our paper using scan strategy of \lrpc\ 
and \remsp\\[1ex]
\hline
\end{tabular}
\label{table:abr} 
\end{table}

We propose two two-pass algorithms for labeling the connected components, \aremsp\ and \nremsp, which are based on \rems\ union-find algorithm
\remsp\ \cite{Patwary2010_RemSP, Dijkstra1976_RemSP} and the scan strategy of \arun\ and
\lrpc\ algorithms.
Since \rems\ union-find is an algorithm which implements immediate parent check test and 
compression technique called {\em Splicing} \cite{Patwary2010_RemSP,
Dijkstra1976_RemSP}, our proposed sequential two-pass algorithm \aremsp\ is $39$\% faster than \lrpc\
and $4$\% faster than \arun.
Another advantage of using \rems\ union-find approach is that its parallel implementation is shown to scale better
with increasing number of processor \cite{Patwary2012_PARemSP}. Parallel \rems\ union-find implementation thus allows us to
process the pixels of the image in any order. Therefore, we propose \paremsp,\ a parallel implementation of our proposed
sequential two-pass CCL algorithm \aremsp. For
scalability, our algorithm in the first pass, divides the image into equal proportions and executes the scan strategy of \arun\ algorithm along with \remsp\
concurrently on each portion of the image. To merge the provisional labels on the image boundary, we use the parallel version
of \remsp \cite{Patwary2012_PARemSP}. Our experiments show
the scalability of \paremsp\ achieving speedups up to $20.1$ using $24$ cores
on shared memory architecture for an image of size $465.2$ MB.
Additionally, the parallel algorithm does not make use of any hardware specific routines, and thus is highly portable.

The remainder of this paper is organized as follows. In section
\ref{sec:related_works}, we provided related work on connected component labeling.
In section \ref{sec:proposed_algorithm}, we propose our sequential two-pass CCL
algorithms \nremsp\ and \aremsp\ and the parallel algorithm \paremsp\ in section
\ref{sec:parallel_algo}.
We present our experimental methodology and results in section
\ref{sec:experiments}. We conclude our work in section
\ref{conclusion}. The abbreviations used in the paper and their brief
description is given in Table \ref{table:abr}.

 \vspace{\sectionSpace}
\section{Related Work}
\label{sec:related_works}

As mentioned in \cite{Suzuki2003_Linear}, there exist different types of CCL
algorithms. Repeated pass or multi pass algorithm repeatedly scans the image
forward and backward alternatively to give labels until no further changes can
be made to the assigned
pixels\cite{Haralick1981_Repeated,Hashizume1990_Algorithm}. 
The algorithm in \cite{Suzuki2003_Linear}, which we call as {\em Suzuki's} algorithm modifies the
conventional multi pass algorithm using one-dimensional table. There exists a
parallel implementation of {\em Suzuki's} algorithm using OpenMP in
\cite{Niknam2010_Parallel}. According to experimental results in
\cite{Niknam2010_Parallel}, the parallel implementation gets maximum speedup of
2.5 on 4 threads.

In any two-pass algorithm, there are two steps in scanning step: $1)$ examining neighbors of current pixel which already
assigned labels to determine label for the current pixel, $2)$ storing label equivalence information to speed up the algorithm. 

The algorithm in \cite{Wu2009_LRPC}, which we refer to as \lrpc, provides two strategies to improve the running time of the algorithm.
First strategy employs a decision tree, which reduces the average number of neighbors accessed by a factor of two.
Second strategy replaces the conventional pointer based union-find algorithm, which is used for storing label equivalence,
by adopting array based union-find algorithm that uses less memory. The
union-find algorithm is implemented using Link by Rank and Path Compression technique. 

The union-find data structure in \cite{He2008_Run} is replaced by a different data structure to process label equivalence
information. In this algorithm, at any point, all provisional labels that are assigned to a connected 
component found thus far during the first scan are combined in a set $S(r)$, where $r$ is the smallest label and is 
referred to as the representative label. The algorithm employs $rtable$ for storing representative label of a set, $next$ to 
find the next element in the set and $tail$ to find the last element of the set.


In another strategy, which we call \arun, the first part of scanning step employs a scanning technique, which processes image two lines 
at a time and process two image pixels at a time \cite{He2012_ARun}. This
algorithm uses the same data structure given in \cite{He2008_Run} for processing label equivalence information. The scanning technique reduces the number lines to be
processed by half thereby improving the speed of the two-pass CCL method. 

In this paper, we provide two different implementations of two-pass CCL algorithm. These two algorithms are different in their
first scan step. In the first implementation called \nremsp, we have used the 
decision tree suggested by the \lrpc\ algorithm for the first part of scanning step but for the second part we have used
\rems\ union-find approach instead of Link by Rank and Path Compression technique. The union-find algorithm
maintains a collection of disjoint sets where each set represents connected elements.
\cite{Patwary2010_RemSP} compares all of the different variations of union-find algorithms over different graph data sets and found that
\rems\ implementation is best among all the variations. Thus in our second implementation, called \aremsp, we process the image lines
two by two as suggested by \cite{He2012_ARun} but for the second step
we use \remsp\ instead of the data structure used by \cite{He2012_ARun}.

We have compared both of our proposed implementations with \lrpc, \run, and
\arun\ algorithms and find that \aremsp\ performs best among all the algorithms.
Finally we have also provided a shared memory parallel implementation of
\aremsp\ called \paremsp\ using OpenMP.
We use the parallel implementation of \remsp\ given in \cite{Patwary2012_PARemSP}. 

 \vspace{\sectionSpace}
\section{Proposed Algorithm}
\label{sec:proposed_algorithm}

Throughout the paper, for an $M \times N$ image, we denote $image(a)$ to denote the pixel value of pixel $a$.
We consider binary images i.e. an image containing two types of pixels:
object pixel and background pixel. Generally, we consider value of object pixel as 1 and value of background pixel as 0. The connected
component labeling problem is to assign a label to each object pixel so that connected object pixels have the same label.
In $2D$ images, there are two ways of defining connectedness: $4$-connectedness and $8$-connectedness. In this paper, we have 
only used $8$-connectedness of a pixel.

\vspace{\sectionSpace}
\subsection{\nremsp\ Algorithm}

In \nremsp,\ we have used the decision tree suggested in \lrpc\ (Figure \ref{dtree})
for scanning and \rems\ union-find algorithm \remsp\ for storing label
equivalence.
The full algorithm for \nremsp\ is given as Algorithm \ref{alg:RemSP}.

\begin{algorithm}[ht!]
\small
{
	\caption{Pseudo-code for \nremsp}
	\label{alg:RemSP}
	\textbf{Input:} $2D$ array $image$ containing the pixel values \\
	\textbf{Output:} $2D$ array $label$ containing the final labels
	\begin{algorithmic}[1]
	\Function{\nremsp}{$image$}
		\State $Scan\_CCLRemSP(image)$ \Comment{Scan Phase of \nremsp}
		\LineComment $count$ is the max label assigned during Scan Phase
		\State $flatten(p,count)$ \Comment{Analysis Phase of \nremsp}
		\For{$row$ in $image$}  \Comment{Labeling Phase of \nremsp}
			\For{$col$ in $row$} \Comment{$e$ is the current pixel to be labeled}
				\State $label(e) \gets p[label(e)]$ 
			\EndFor
		\EndFor
	\EndFunction
	\end{algorithmic}
}	
\end{algorithm}

\begin{figure}[h]
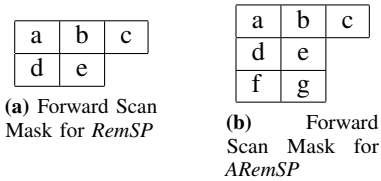

\centering

\subfloat[Forward Scan Mask for \em{RemSP}]{
\begin{tabular}{ | l | c | r | }
 \hline
 a & b & c \\ \hline
 d & e & \multicolumn{1}{r}{} \\ \cline{1-2}
\end{tabular}
\label{fscan1}
}
\hspace{0.8cm}
\subfloat[Forward Scan Mask for \em{ARemSP}]{
\begin{tabular}{ | l | c | r | }
 \hline
 a & b & c \\ \hline
 d & e & \multicolumn{1}{r}{} \\ \cline{1-2}
 f & g & \multicolumn{1}{r}{} \\ \cline{1-2}
\end{tabular}
\label{fscan2}
}
\caption{Forward Scan Mask}
\label{fscan}
\end{figure}

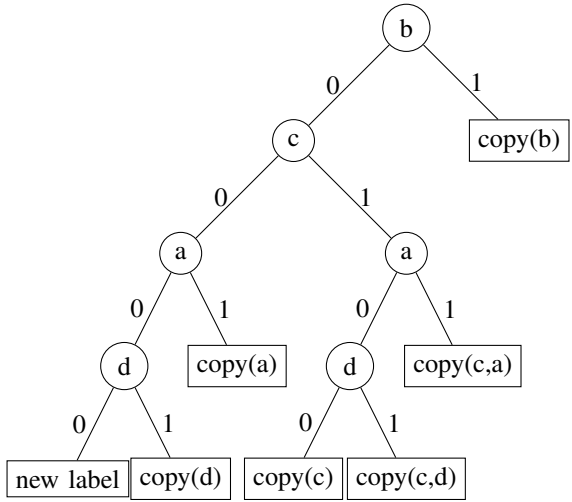
\begin{figure}[h]
\centering
\begin{tikzpicture}[level distance=1.5cm,
  level 1/.style={sibling distance=3 cm},
  level 2/.style={sibling distance=3 cm},
  level 3/.style={sibling distance=1.5cm},
  level 4/.style={sibling distance=1.5cm}]
  \node[circle,draw] {b}
   child 
   {
    	node[circle,draw] {c}
      	child 
      	{
      	    	node[circle,draw] {a}
      	    	child
      	    	{
      	    		node[circle,draw]{d}
      	    		child
      	    		{
      	    			node[rectangle,draw]{new label}
      	    			edge from parent node[left,draw=none] {0}
      	    		}
      	    		child
      	    		{
      	    			node[rectangle,draw]{copy(d)}
      	    			edge from parent node[right,draw=none] {1}
      	    		}
      	    		edge from parent node[left,draw=none] {0}
      	    	}
      	    	child
      	    	{
      	    		node[rectangle,draw]{copy(a)}	
      	    		edge from parent node[right,draw=none] {1}
      	    	}
      	    	edge from parent node[left,draw=none] {0}
      	}
      	child 
      	{
      		node[circle,draw] {a}
      		child
      		{
      			node[circle,draw]{d}
      	    		child
      	    		{
      	    			node[rectangle,draw]{copy(c)}
      	    			edge from parent node[left,draw=none] {0}
      	    		}
      	    		child
      	    		{
      	    			node[rectangle,draw]{copy(c,d)}
      	    			edge from parent node[right,draw=none] {1}
      	    		}
      	    		edge from parent node[left,draw=none] {0}
      		}
      		child
      		{
      			node[rectangle,draw]{copy(c,a)}
      			edge from parent node[right,draw=none] {1}
      		}
      		edge from parent node[right,draw=none] {1}
      	}
      	edge from parent node[left,draw=none] {0}
    }
    child 
    {
    	node[rectangle,draw] {copy(b)}
    	edge from parent node[right,draw=none] {1}
    };
\end{tikzpicture}
\caption{Decision tree suggested in \em{CCLLRPC}\cite{Wu2009_LRPC}}
\label{dtree}
\end{figure}

In the scan step of \nremsp, we process image lines one by one using the
forward scan mask as shown in Figure \ref{fscan1}. We have used the decision tree proposed by \cite{Wu2009_LRPC} for 
determining the provisional label of current pixel $e$ as we can reduce the number of neighbors using decision tree. Instead of
examining all four neighbors of pixel, say $e$, i.e. $a$, $b$, $c$ and $d$, we only
examine the neighbors according to a decision tree as shown in Figure
\ref{dtree}\cite{Wu2009_LRPC}.
 Let $label$ denote the $2D$ array storing the labels and let $p$ denote equivalence array 
 then according to \lrpc\ algorithm,
three functions used by this decision tree are defined as follows:

$1)$. The one-argument copy function, copy(a), contains one statement:
					$label(e) = p(label(a))$\\
$2)$. The two-argument copy function, copy(c,a), contains one statements:
				$label(e) = merge(p, label(c), label(a))$\\
$3)$. The new label function sets $count$ as $label(e)$, appends $count$ to array $p$, and increments $count$ by $1$.

The implementation of {\em Scan\_CCLRemSP} is given as Algorithm
\ref{alg:RemSP-I}. However, the implementation of \merge\ operation in our
proposed algorithm \remsp\ is different from that of in \lrpc.\
We have used the implementation of union-find proposed by \rems\ 
for merge operation\cite{Patwary2010_RemSP, Dijkstra1976_RemSP}.\ \rems\ integrates the {\em union}
operation with a compression technique known as Splicing $(SP)$. 

In the \merge\ algorithm, if $x$ and $y$ are the nodes to be merged then we set $root_x$ to $x$ and $root_y$ to $y$. 
When $root_x$ is to be moved to $p(root_x)$, firstly $p(root_x)$ is
stored in a temporary variable $z$ then $p(root_x)$ is set to $p(root_y)$,
making the subtree rooted at $root_x$ a sibling of $root_y$ and finally
$root_x$ is set to $z$. The algorithm for \merge\ is given as Algorithm
\ref{alg:merge}. After the first step, we carry out the analysis phase using the \flatten\ algorithm. In the \flatten\ algorithm, we give smallest equivalent label of every connected component to all the pixels belonging to that connected component. The algorithm also
generates consecutive labels. The algorithm for \flatten\ is given as Algorithm
\ref{alg:flatten}. 
\begin{algorithm}[ht!]
\small
{
	\caption{Pseudo-code for merge\cite{Patwary2010_RemSP}}
	\label{alg:merge}
	\textbf{Input:} $1D$ array $p$ and two nodes $x$ and $y$ \\
	\textbf{Output:} The root of united tree 
	\begin{algorithmic}[1]
	\Function{merge}{$p$,$x$,$y$} 
		\State $root_x \gets x, root_y \gets y$
		\While { $p[root_x] \neq p[root_y]$ }
			\If {$p[root_x] > p[root_y]$}
				\If { $root_x = p[root_x]$ }
					\State $p[root_x] \gets p[root_y]$
					\State \Return{$p[root_x]$}
				\EndIf
				\State $z \gets p[root_x], p[root_x] \gets p[root_y], root_x \gets z$
			\Else
				\If { $root_y = p[root_y]$ }
					\State $p[root_y] \gets p[root_x]$
					\State \Return{$p[root_x]$}
				\EndIf
				\State $z \gets p[root_y], p[root_y] \gets p[root_x], root_y \gets z$
			\EndIf
		\EndWhile
		\State \Return{$p[root_x]$}
	\EndFunction
	\end{algorithmic}	
}
\end{algorithm}

\begin{algorithm}[H]
\small
{
	\caption{Pseudo-code for flatten \cite{Wu2009_LRPC}}
	\label{alg:flatten}
	\textbf{InOut:} $1D$ array $p$ containing the equivalence info \\
	\textbf{Input:} Max value of provisional label $count$
	\begin{algorithmic}[1]
	\Function{flatten}{$p$,$count$}
		\State $k \gets 1$
		\For{$i=1$ to $count$}
			\If{$p[i] < i$}
				\State $p[i] \gets p[p[i]]$
			\Else
				\State $p[i] \gets k$
				\State $k++$
			\EndIf
		\EndFor
	\EndFunction
	\end{algorithmic}	
}
\end{algorithm}

\subsection{\aremsp\ Algorithm}

In \aremsp,\ we have used the decision tree suggested in \arun\ 
for scanning and \rems\ union-find algorithm for storing label equivalence. The
full algorithm for \aremsp\ is given as Algorithm \ref{alg:ARemSP}.

In the first scan step of \aremsp,\ we process an image two lines at a time and
two pixels at a time using the mask shown in Figure \ref{fscan2}, which is
suggested in \cite{He2012_ARun}.
\clearpage
\begin{algorithm}[H]
\small
{
	\caption{Pseudo-code for \nremsp\ Scan Phase}
	\label{alg:RemSP-I}
	\textbf{Input:} $2D$ array $image$ containing the pixel values \\
	\textbf{InOut:} $2D$ array $label$ containing the provisional labels and $1D$ array $p$ containing the equivalence info\\
	\textbf{Output:} maximum value of provisional label in $count$
	\begin{algorithmic}[1]
	\Function{Scan\_CCLRemSP}{$image$}
		\For{$row$ in $image$}
			\For{$col$ in $row$}
				\If{$image(e) = 1$}
					\If{$image(b) = 1$}
						\State $copy(b)$
					\Else
						\If{$image(c) = 1$}
							\If{$image(a) = 1$}
								\State $copy(c,a)$
							\Else
								\If{$image(d) = 1$}
									\State $copy(c,d)$
								\Else
									\State $copy(c)$
								\EndIf
							\EndIf
						\Else
							\If{$image(a) = 1$}
								\State $copy(a)$
							\Else
								\If{$image(d) = 1$}
									\State $copy(d)$
								\Else
									\State {\em new label}
								\EndIf
							\EndIf
						\EndIf
					\EndIf
				\EndIf
			\EndFor
		\EndFor
		\State \Return {$count$}
	\EndFunction
	\end{algorithmic}
}	
\end{algorithm}

\begin{algorithm}[H]
\small
{
	\caption{Pseudo-code for ARemSP}
	\label{alg:ARemSP}
	\textbf{Input:} $2D$ array $image$ containing the pixel values \\
	\textbf{Output:} $2D$ array $label$ containing the final labels
	\begin{algorithmic}[1]
	\Function{ARemSP}{$image$}
		\State $Scan\_ARemSP(image)$ \Comment{Scan Phase of RemSP}
		\LineComment $count$ is the max label assigned during Scan Phase
		\State $flatten(p,count)$ \Comment{Analysis Phase of RemSP}
		\For{$row$ in $image$}  \Comment{Labeling Phase of RemSP}
			\For{$col$ in $row$} \Comment{$e$ is the current pixel to be labeled}
				\State $label(e) \gets p[label(e)]$
			\EndFor
		\EndFor		
	\EndFunction
	\end{algorithmic}	
}
\end{algorithm}

We assign the label to both $e$ and $g$ simultaneously. If both $e$ and $g$ are
background pixels, then nothing needs to be done. If $e$ is a foreground pixel and there is no foreground pixel in the mask, we assign a 
new provisional label to $e$ and if $g$ is a foreground pixel, we will assign
the label of $e$ to $g$. If there are foreground pixels in the mask, then we assign $e$ any label assigned to 
foreground pixels. In this case, if there is only one connected component in the mask then there is 
no need for label equivalence. Otherwise, if there are more than one connected components in the mask and as 
they are connected to $e$, all the labels of the connected components are
equivalent labels and hence need to be merged. For all the cases, one can refer
to \cite{He2012_ARun}.
However, our implementation for the merge operation is different from \cite{He2012_ARun}.
We use the implementation of union-find proposed by {\em Rem} \cite{Patwary2010_RemSP, Dijkstra1976_RemSP} for the merge operation in
\aremsp.\ Similar to \nremsp,\ we use \flatten\ for analysis phase
and generating consecutive labels. The implementation of {\em Scan\_ARemSP} is given as Algorithm
\ref{alg:ARemSP-I}.

\begin{algorithm}[ht!]
\small
{
	\caption{Pseudo-code for ARemSP Scan Phase}
	\label{alg:ARemSP-I}
	\textbf{Input:} $2D$ array $image$ containing the pixel values \\
	\textbf{InOut:} $2D$ array $label$ containing the provisional labels and $1D$ array $p$ containing the equivalence info\\
	\textbf{Output:} maximum value of
	provisional label in $count$
	\begin{algorithmic}[1]
	\Function{Scan\_ARemSP}{$image$}
		\For{$row$ in $image$}
			\For{$col$ in $row$}
				\If{$image(e) = 1$}
					\If{$image(d) = 0$}
						\If{$image(b) = 1$}
							\State $label(e) \gets label(b)$
							\If{$image(f) = 1$}
								\State $merge(p,label(e),label(f))$ 
							\EndIf
						\Else
							\If{$image(f) = 1$}
								\State $label(e) \gets label(f)$
								\If{$image(a) = 1$}
									\State $merge(p,label(a))$
								\EndIf
								\If{$image(c) = 1$}
									\State $merge(p,label(e),label(c))$
								\EndIf
							\Else
								\If{$image(a) = 1$}
									\State $label(e) \gets label(a)$
									\If{$image(c) = 1$}
										\State $merge(p,label(e),label(c))$
									\EndIf
								\Else
									\If{$image(c) = 1$}
										\State $label(e) \gets label(c)$
									\Else
										\State $label(e) \gets count,$
										\State $p[count] \gets count,$
										\State $count++$
									\EndIf
								\EndIf
							\EndIf
						\EndIf
					\Else
						\State $label(e) = label(d)$
						\If{$image(b) = 0$}
							\If{$image(c) = 1$}
								\State $merge(p,label(e),label(c))$
							\EndIf
						\EndIf
					\EndIf
					\If{$image(g) = 1$}
						\State $label(g) \gets label(e)$
					\EndIf
				\Else
					\If{$image(g) = 1$}
						\If{$image(d) = 1$}
							\State $label(g) \gets label(d)$
						\Else
							\If{$image(f) = 1$}
								\State $label(g) \gets label(f)$
							\Else
								\State $label(e) \gets count,$
								\State $p[count] \gets count,$
								\State $count++$
							\EndIf
						\EndIf
					\EndIf
				\EndIf
			\EndFor
		\EndFor
		\State \Return {$count$}
	\EndFunction
	\end{algorithmic}	
}
\end{algorithm}

 \vspace{\sectionSpace}
\section{Parallelizing \aremsp\ Algorithm}
\label{sec:parallel_algo}

We now describe the parallel implementation of \aremsp\ algorithm 
on a shared memory system. We make the assumption about memory
model as stated in {\em OpenMP} regarding the atomic directive. We assume that
memory read/write operations are atomic and any operations issued concurrently by different processors will be executed in some unknown sequential order if
no ordering constructs are being used. However, two dependent operations issued
by the same processor will always be applied in the same order as they are
issued.
\begin{algorithm}[ht!]
\small
{
	\caption{Pseudo-code for PARemSP}
	\label{alg:PARemSP}
	\textbf{Input:} $2D$ array $image$ containing the pixel values \\
	\textbf{Output:} $2D$ array $label$ containing the final labels
	\begin{algorithmic}[1]
	\Function{PARemSP}{$image$}
		\State $num iter \gets row/2 $ \Comment{As we are processing $2$ rows at a time}
		\State \# pragma omp parallel
			\State $chunk \gets numiter/number of threads $ 
			\State $size \gets 2 \times chunk$
			\State $start \gets $ start index of the thread
			\State $count \gets start \times col$
			\State \# pragma omp for
				\State $Scan\_ARemSP(image)$
			\State \# pragma omp for
				\For {$i=size$ to $row-1$}
					  \For{$col$ in $row$}
					  	\If{$label(e) \neq 0$}
					  		\If{$label(b) \neq 0$}
					  			\State $merger(p,label(e),label(b))$
					  		\Else
					  			\If{$label(a) \neq 0$}
									\State $merger(p,label(e),label(a))$
								\EndIf
								\If{$label(c) \neq 0$}
									\State $merger(p,label(e),label(c))$
								\EndIf
							\EndIf
						\EndIf
					\EndFor
					\State $i \gets i + size$
				\EndFor		
		\State $flatten(p,count)$ 
		\For{$row$ in $image$}  
			\For{$col$ in $row$}
				\State $label(e) \gets p[label(e)]$
			\EndFor
		\EndFor		
	\EndFunction
	\end{algorithmic}	
}
\end{algorithm}

\begin{algorithm}[ht!]
\small
{
	\caption{Pseudo-code for merger\cite{Patwary2012_PARemSP}}
	\label{alg:merger}
	\textbf{Input:} $1D$ array $p$ and two nodes $x$ and $y$ \\
	\textbf{Output:} The root of united tree 
	\begin{algorithmic}[1]
	\Function{merger}{$p$,$x$,$y$} 
		\State $root_x \gets x, root_y \gets y$
		\While { $p[root_x] \neq p[root_y]$ }
			\If {$p[root_x] > p[root_y]$}
				\If{$root_x = p[root_x]$}
					\State $success \gets 0$
					\State $omp\_set\_lock(\&(lock\_array[root_x]))$
					\If{$root_x = p[root_x]$}
						\State $p[root_x] \gets p[root_y]$
						\State $success \gets 1$
					\EndIf
					\State $omp\_unset\_lock(\&(lock\_array[root_x]))$
					\If{$success =1$}  
						\State break
					\EndIf
				\EndIf
				\State $z \gets p[root_x], p[root_x] \gets p[root_y], root_x \gets z$
			\Else
				\If{$root_y = p[root_y]$}
					\State $success \gets 0$
					\State $omp\_set\_lock(\&(lock\_array[root_y]))$
					\If{$root = p[root_y]$}
						\State $p[root_y] \gets p[root_x]$
						\State $success \gets 1$
					\EndIf
					\State $omp\_unset\_lock(\&(lock\_array[root_y]))$
					\If{$success =1$}  
						\State break
					\EndIf
				\EndIf
				\State $z \gets p[root_y], p[root_y] \gets p[root_x], root_y \gets z$
			\EndIf
		\EndWhile
		\State \Return{$p[root_x]$}
	\EndFunction
	\end{algorithmic}	
}
\end{algorithm}

In \paremsp, the image is
divided row-wise into chunks of equal size and given to the threads. In the first step,
each thread runs {\em Scan Phase} of \aremsp\ on it's chunk simultaneously. We
initialize the label to the start index of the thread for every thread so that no 
two pixels in the image have the same label after the first step. After the first step, 
each pixel is given a provisional label. Next, the pixels at the boundary of
each chunk need to be merged to get the final labels. In the second step, we merge the boundary pixels using parallel 
implementation of Rem's Algorithm \cite{Patwary2012_PARemSP} which we call as
\merger.\ In \merger,\ if a thread wants to perform merging, it will first
acquire the necessary lock. Once it gets the lock, it will check whether the
node is still a root node. If yes, then the thread
will set the parent pointer and release the lock. On the other
hand if some other processor has altered the parent pointer so
that the node is no longer a root, the processor will release
the lock and continue executing the algorithm from its current
position. For complete reference, one can refer \cite{Patwary2012_PARemSP}. We
implement the parallel algorithm using OpenMP directives {\em pragma omp parallel} and {\em pragma omp for}. 
The pseudo code of \merger\ is given as Algorithm \ref{alg:merger}. 
The pseudo code of \paremsp\ is given as Algorithm
\ref{alg:PARemSP}.

 \vspace{\sectionSpace}
\section{Experiments}
\label{sec:experiments}
For the experiments we used a computing node of Hopper, a Cray XE6 distributed memory parallel computer. 
The node has 2 twelve-core AMD ‘MagnyCours’ 2.1-GHz processors and 32 GB DDR3 1333-MHz memory. 
Each core has its own L1 and L2 caches, with 64 KB and 512 KB, respectively. 
One 6-MB L3 cache is shared between 6 cores on the MagnyCours processor. 
All algorithms were implemented in C using OpenMP and compiled with gcc.

Our test data set consists of four types of image data set: Texture, Aerial,
Miscellaneous and NLCD. First three data sets are taken from the image database of the University of 
Southern
California.\footnote{\url{http://sipi.usc.edu/database/}} 
The fourth data set is taken from US National Cover Database
$2006$.\footnote{\url{http://dx.doi.org/10.1016/j.cageo.2013.05.014}} All of the
images are converted to binary images by MATLAB using $im2bw(level)$ function with level value as $0.5$. The function 
converts the grayscale image to a binary image by replacing all pixels in the input image with luminance greater than 
0.5 with the value 1 (white) and replaces all other pixels with the value 0 (black). If the input image is not a grayscale image, 
$im2bw$ converts the input image to grayscale, and then converts this grayscale image to binary(Figure \ref{fig:example}). However, note that our
algorithm can be easily extended to gray scale images.

\begin{figure}[ht!]
	\begin{center}
	\subfloat[{Original Image}]
        {
                \label{fig:color}
                \includegraphics[width=\scaleLarge]{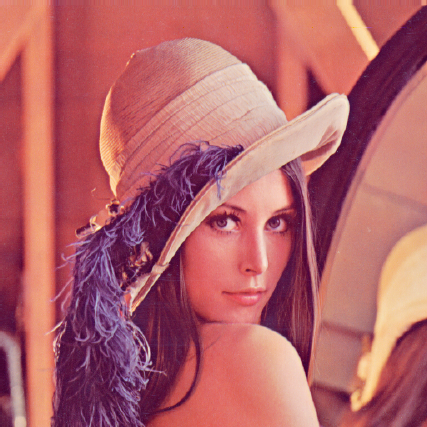}
        }
	\hspace{3mm}
	\subfloat[{Binary Image}]
        {
                \label{fig:bw}
                \includegraphics[width=\scaleLarge]{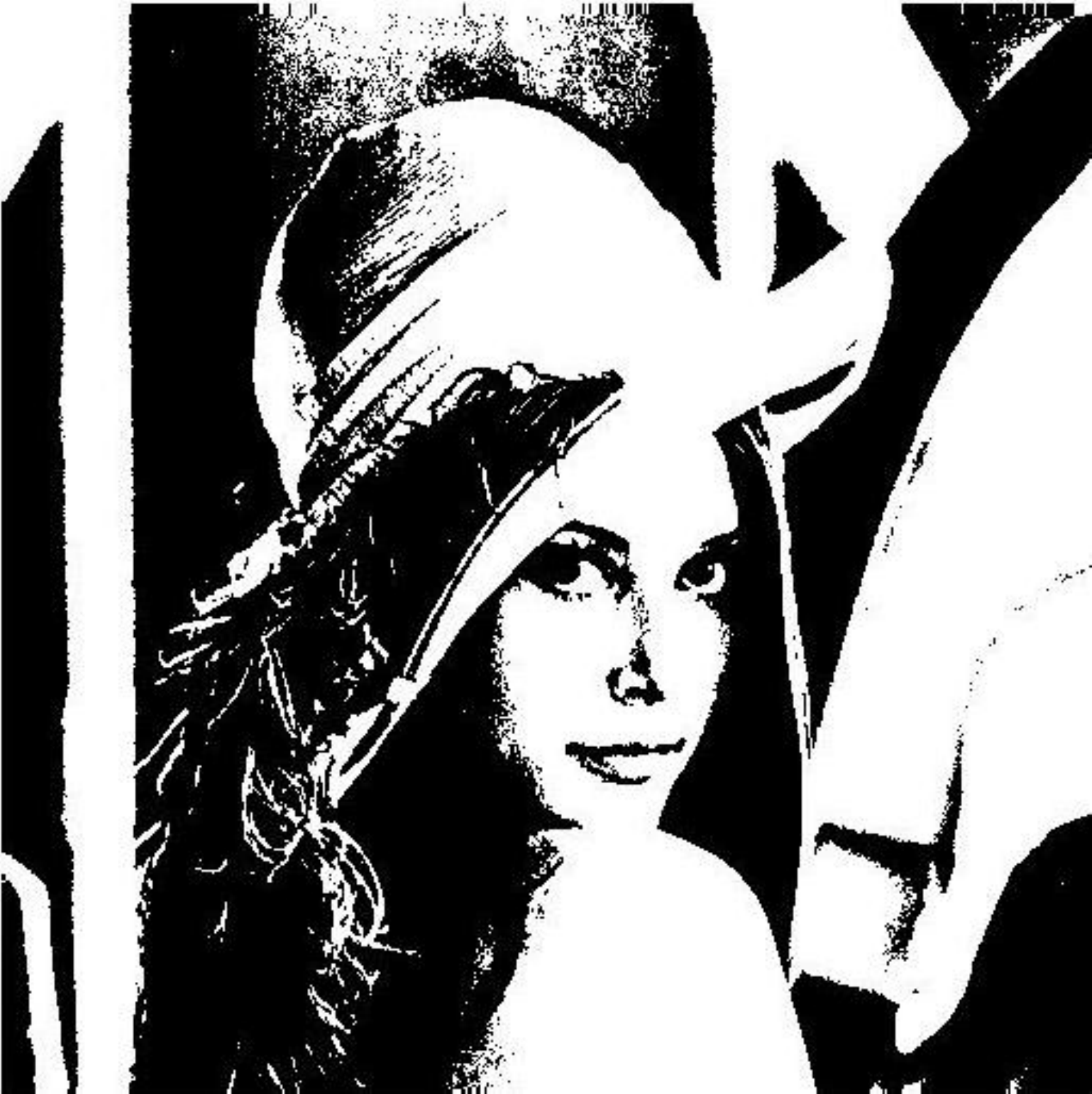}
        }
	\end{center}
	\caption{Example of color image to binary image}
	\label{fig:example}
\end{figure}
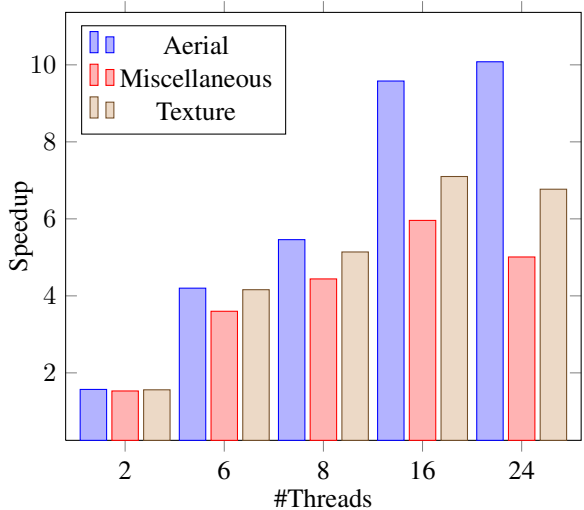
\begin{figure}[h]
\centering
\begin{tikzpicture}
\begin{axis}[
ybar,
enlargelimits=0.15,
legend pos = north west,
xlabel=\#Threads,
ylabel=Speedup,
every axis y label/.style=
{at={(ticklabel cs:0.5)},rotate=90,anchor=center},
symbolic x coords={2,6,8,16,24},
xtick=data,
]
\addplot
coordinates {
(2,1.57)
(6,4.20)
(8,5.46)
(16,9.58)
(24,10.08)

};
\addplot
coordinates {
(2,1.53)
(6,3.60)
(8,4.44)
(16,5.96)
(24,5.01)
};
\addplot
coordinates {
(2,1.56)
(6,4.16)
(8,5.14)
(16,7.10)
(24,6.77)
};

\legend{Aerial,Miscellaneous,Texture}
\end{axis}
\end{tikzpicture}
\caption{Speedup for different images and different numbers of threads for
Aerial, Texture \& Miscellaneous data set}
\label{fig:bar}
\end{figure}
Texture, Aerial and Miscellaneous data set contain images of size $1$ MB or less.
NCLD data set contains images of size bigger than $12$ MB. The biggest image in the data set is $465.20$ MB.

Firstly, we performed the experiment over all the sequential algorithms. The
experimental results are shown in Table \ref{table:seq}. In the table, we have
shown the minimum, maximum and average execution time of all the four data sets.
The execution time of \aremsp\ is lowest among all
the sequential algorithms. Thus \aremsp\ is best among all the sequential
algorithms.

\begin{figure*}[h!]
\begin{center}
\subfloat[local]{
\begin{tikzpicture}
\begin{axis}[
legend columns= -1,
legend
entries={image\_1, image\_2,
image\_3, image\_4, image\_5, image\_6, ideal}, legend to name=named, 
xlabel=\# Threads,
ylabel=Speedup,
every axis y label/.style=
{at={(ticklabel cs:0.5)},rotate=90,anchor=near ticklabel},
]

\addplot+[sharp plot] coordinates{
(2,1.54)
(6,4.82)
(12,8.98)
(18,13.04)
(24,16.70)
};
\addplot+[sharp plot] coordinates {
(2,1.58)
(6,5.22)
(12,9.7)
(18,13.65)
(24,16.76)
};
\addplot+[sharp plot] coordinates{
(2,1.45)
(6,5.12)
(12,9.53)
(18,13.77)
(24,17.74)
};
\addplot+[sharp plot] coordinates{
(2,1.53)
(6,5.30)
(12,10.30)
(18,14.97)
(24,19.18)
};
\addplot+[sharp plot] coordinates{
(2,1.64)
(6,5.47)
(12,10.38)
(18,15.12)
(24,19.77)
};

\addplot+[sharp plot] coordinates{
(2,1.7)
(6,5.54)
(12,10.74)
(18,15.45)
(24,20.10)
};
\addplot+[sharp plot] coordinates{
(2,2)
(6,6)
(12,12)
(18,18)
(24,24)
};
\end{axis}
\end{tikzpicture}
\label{fig:line}
}
\subfloat[local + merge]{
\begin{tikzpicture}
\begin{axis}[
xlabel=\# Threads,
ylabel=Speedup,
every axis y label/.style=
{at={(ticklabel cs:0.5)},rotate=90,anchor=near ticklabel},
]
\addplot+[sharp plot] coordinates{
(2,1.48)
(6,4.80)
(12,8.91)
(18,12.86)
(24,16.30)
};

\addplot+[sharp plot] coordinates {
(2,1.54)
(6,5.18)
(12,9.52)
(18,13.06)
(24,15.64)
};
\addplot+[sharp plot] coordinates{
(2,1.40)
(6,5.11)
(12,9.49)
(18,13.69)
(24,17.59)
};

\addplot+[sharp plot] coordinates{
(2,1.49)
(6,5.29)
(12,10.23)
(18,14.75)
(24,18.73)
};

\addplot+[sharp plot] coordinates{
(2,1.60)
(6,5.46)
(12,10.34)
(18,15.02)
(24,19.58)
};

\addplot+[sharp plot] coordinates{
(2,1.55)
(6,5.53)
(12,10.72)
(18,15.42)
(24,20.03)
};
\addplot+[sharp plot] coordinates{
(2,2)
(6,6)
(12,12)
(18,18)
(24,24)
};
\end{axis}
\end{tikzpicture}
\label{fig:linet}
}
\label{speed}
\ref{named}
\caption{Speedup for different images and different numbers of threads for NLCD
data set}
\end{center}
\end{figure*}
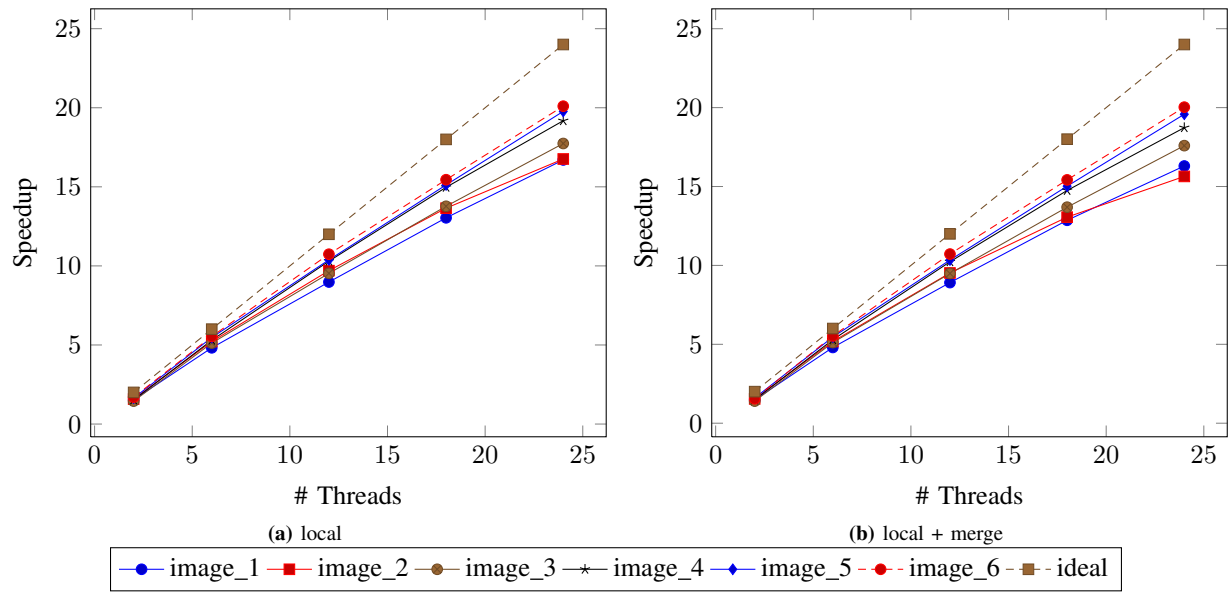

Next, we show our results for the parallel algorithm \paremsp\ over all the images.
We have shown the speedup for data sets (except NCLD image data set) in Figure \ref{fig:bar}.
The minimum, maximum and average execution time of \paremsp\ for all the datasets is also shown in Table \ref{table:par}.
We get a maximum speedup of $10$ in this case as the images are $1$ MB or less in size. The speedup also decreases in some cases as the
number of threads increases.
This case occurs when the image size is small. As the number of threads increase, each threads has less work, and therefore the thread creation and termination overhead will affect the performance.
Figure \ref{fig:line}-\ref{fig:linet} shows the speedup of the algorithm for 
NCLD image data set. The size of the images are given in Table \ref{table:imag_size}.
We get a maximum speedup of $20.1$ on $24$ cores for image of size $465.20$ MB.
Figure \ref{fig:line} shows the speedup for {\em Phase-$I$} of \paremsp\ i.e. 
the local computation and Figure \ref{fig:linet} shows the overall speedup (i.e.
local + merge). We can see that there is not significant difference between both speedups, implying that merge operation
does not have a significant overhead. 
Also Figure 5 shows, speedup increases with image size.

Therefore, our parallel implementation is able to achieve near linear
speed for large data sets.

\begin{table}[h!]
\caption{Comparison of various execution times[{\em msec}] for sequential algorithms}
\centering
\begin{tabular}{c c c c c c} 
\hline\hline
Image type &  & CCLLRPC & CCLRemSP & ARun & ARemSP\\ [0.25ex] 
\hline 
Aerial & Min & 2.5 & 2.48 & 1.98 & 1.95 \\ 
 & Average & 13.68 & 13.25 & 11.90 & 11.86\\
 & Max & 86.64 & 80.90 & 72.92 & 70.17 \\
[0.25ex]
 \hline
Texture & Min &2.07 & 2.06 & 1.58 & 1.53 \\
 & Average & 8.42 & 8.20 & 7.32 & 7.27 \\
 & Max & 16.86 & 16.18 & 14.81 & 14.47\\
[0.25ex]
 \hline
Misc & Min & 0.50 & 0.49 & 0.36 & 0.36\\
 & Average & 3.28 & 3.21 & 2.75 & 2.74 \\
 & Max &12.96 & 12.81 & 11.30 & 11.20\\
 [0.25ex]
\hline
NLCD & Min & 4.61 & 4.46	& 3.77 & 3.75 \\
& Average & 307.66 & 299.55 & 244.88 & 242.59 \\
& Max & 1307.27	& 1273.82 & 1036.52 & 1021.45
\\[0.25ex]
\hline
\end{tabular}
\label{table:seq} 
\end{table}

\begin{table}[h]
\caption{Images and their sizes [in {\em MB}]}
\centering
\begin{tabular}{c c} 
\hline\hline
Image name & Size\\ [0.5ex] 
\hline 
image\_1 & $12$\\
image\_2 & $33$ \\
image\_3 & $37.31$ \\
image\_4 & $116.30$ \\
image\_5 & $132.03$\\ 
image\_6 & $465.20$\\
[1ex]
\hline
\end{tabular}
\label{table:imag_size} 
\end{table}

\begin{table}[ht!]
\caption{Execution time [{\em msec}] of \paremsp\ algorithm for various \#
threads}
\centering
\begin{tabular}{c c c c c c} 
\hline\hline
Image type &  & 2 & 6 & 16 & 24\\ [0.5ex] 
\hline 
Aerial & Min & 1.39 & 0.84 & 1.02 & 1.38 \\ 
 & Average & 7.92	& 3.03	& 1.87	& 2.15
\\
 & Max & 46.86 &	16.72 & 	7.32 &	6.97
 \\
[1ex]
 \hline
Texture & Min & 1.09 & 0.62 & 	0.93 &	1.36
 \\
 & Average & 4.91 &	1.99 &	1.45	& 1.82
 \\
 & Max & 9.75 &	3.56 &	2.11 &	2.34
\\
[1ex]
 \hline
Miscellaneous & Min & 0.36	& 0.36	& 0.79	& 1.18
\\
 & Average & 1.99	& 0.97	& 1.05	& 1.46
 \\
 & Max &7.96	& 3.24	& 1.91	& 2.27
\\
 [1ex]
\hline
NLCD & Min & 2.52& 1.16&	1.32&	1.67
\\
& Average & 162.86&	58.50&	20.20&	13.47
 \\
& Max & 676.41	&184.71	&78.33	&51.00
 \\[1ex]
\hline
\end{tabular}
\label{table:par} 
\end{table}

 \vspace{\sectionSpace}
\section{Conclusion}
\label{conclusion}
In this paper, we presented two sequential CCL algorithms \nremsp\ and \aremsp\
which are based on union-find technique of \rems\ algorithm and scan strategies 
of \arun\ and \lrpc\ algorithms. \nremsp\ algorithm uses the scan strategy of
\lrpc\ algorithm whereas \aremsp\ uses the scan strategy of \arun\ algorithm.
Based on the experiments, we found out that \aremsp\ outperforms all the
other sequential algorithms. We also implement a portable parallel
implementation of \aremsp\ for shared memory computers with standard OpenMP
directives.
Our proposed algorithm, \paremsp,\ divides the image into equal proportions and
executes the scan. 
To merge the provisional labels on the image boundary, we use the parallel version of \rems\ algorithm. 
Our experimental results conducted on a shared memory computer show scalable performance, achieving speedups up
to a factor of $20.1$ when using $24$ cores on data set of size $465.20$ MB. Thus, our parallel algorithm achieves linear scaling for large
fixed problem size while the number of processing elements are increased. 

 \vspace{\sectionSpace}
\section*{Acknowledgment}
This work is supported in part by the following grants: NSF awards CCF-0833131, CNS-0830927, 
IIS-0905205, CCF-0938000, CCF-1029166, ACI-1144061, and IIS-1343639; DOE awards DE-FG02-08ER25848, 
DE-SC0001283, DE-SC0005309, DESC0005340, and DESC0007456; AFOSR award FA9550-12-1-0458.


\renewcommand{\baselinestretch}{1}
\normalsize
\addcontentsline{toc}{chapter}{\numberline{}{Bibliography}}%
\small{
\bibliography{main}
\bibliographystyle{unsrt}
}

\end{document}